\documentstyle[aps,prl,epsf]{revtex}

\def\ket#1{#1}

\begin{document} 
\title{Coherent population transfer beyond the adiabatic limit: 
generalized matched 
pulses and higher-order trapping states} 
\author{M. Fleischhauer$^1$, R. Unanyan$^{2,3}$, B.W. Shore$^{2,4}$, 
and K. Bergmann$^2$} 
\address{$^1$ Sektion Physik, Ludwig-Maximilians Universit\"at M\"unchen,\\ 
D-80333 M\"unchen, Germany} 
\address{$^2$ Fachbereich Physik, Universit\"at Kaiserslautern,\\ 
D-67663 Kaiserslautern, Germany} 
\address{$^3$ Inst. for Physical Research of the Armenian Nat. Academy of\\ 
Sciences,\\ 
Ashtarak-2, 378410, Armenia} 
\address{$^4$ Lawrence Livermore National Laboratory, Livermore CA 94550,\\ 
USA} 
\date{\today} 
\maketitle 
 
\begin{abstract} 
We show that the physical mechanism of population transfer in a 3-level 
system with a closed loop of coherent couplings (loop-STIRAP) is not 
equivalent to an 
adiabatic rotation of the dark-state of the Hamiltonian but coresponds to 
a rotation of 
a {\it higher-order trapping state} in a generalized adiabatic basis. The 
concept of generalized adiabatic basis sets is used as a constructive tool 
to design pulse sequences for stimulated Raman adiabatic passage (STIRAP) 
which give maximum population transfer also under conditions when the usual 
condition of adiabaticty is only poorly fulfilled. Under certain conditions 
for the pulses (generalized matched pulses) there exists a higher-order 
trapping state, which is an exact constant of motion and analytic solutions 
for the atomic dynamics can be derived. 
\end{abstract} 
 
\pacs{1234567890} 
 
% INITIALIZE - DONT CHANGe 
 
%%%%%%%%%%%%%%%%%%%%%%%%%%%%%%%%%%%%%%%%%%%%%%%%%%%%%%%%%%%%%%%%%%%%%%%%%%%%%% 
 
%%%%%%%%%%%%%%%%%%%%%%%%%%%%%%%%%%%%%%%%%%%%%%%%%%%%%%%%%%%%%%%%%%%%%%%%%%%%%% 
 
\tighten 
 
%%%%%%%%%%%%%%%%%%%%%%%%%%%%%%%%%%%%%%%%%%%%%%%%%%%%%%%%%%%%%%%%%%%%%%%%%%%%%% 
 
%%%%%%%%%%%%%%%%%%%%%%%%%%%%%%%%%%%%%%%%%%%%%%%%%%%%%%%%%%%%%%%%%%%%%%%%%%%%%% 
 
%\narrowtext 
 
%%%%%%%%%%%%%%%%%%%%%%%%%%%%%%%%%%%%%%%%%%%%%%%%%%%%%%%%%%%%%%%%%%%%%%%%%%%%%% 
 
\section{introduction} 
 
%%%%%%%%%%%%%%%%%%%%%%%%%%%%%%%%%%%%%%%%%%%%%%%%%%%%%%%%%%%%%%%%%%%%%%%%%%%%%% 
 
The transfer of population in a multi-level atomic system from an initial to 
a target quantum state in a fast and effective way is currently a problem of 
practical importance as well as of substantial theoretical interest. If 
there is a dipole allowed transition between an initial and a target state, 
one 
can achieve the desired transfer by using either a constant-frequency $\pi$% 
-pulse tuned to resonance, or an adiabatic process based on a swept 
carrier-frequency. Since a dipole-allowed transition implies radiative 
decay, one is however often interested in systems with two metastable states 
without a direct electric-dipole coupling. Whereas an extension of the 
two-state $\pi$-pulse approach to multistate excitation is possible, these 
techniques require careful control of the pulse areas. Adiabatic processes 
do not require such precise control, if the time-evolution is slow (meaning, 
generally, large pulse areas). In a three-state Raman-transition system, for 
example, it is possible to achieve adiabatic passage with the use of two 
constant-frequency pulses suitably delayed (counterintuitive order) 
\cite{Oreg}. The 
process of this stimulated Raman adiabatic passage (STIRAP) 
\cite{STIRAP,STIRAP2} 
can be  represented by a slow rotation of a decoupled eigenstate of the H
amiltonian (dark 
state) \cite{Arimondo}. 
 
The disadvantage of STIRAP is the requirement for large pulse areas: to 
ensure adiabatic time evolution the effective average Rabi-frequency of the 
pulses must be large compared to the radiative decay rates of the 
intermediate level(s). Non-adiabatic corrections and the associated diabatic 
losses \cite{Elk95,Stenholm96,Fleischhauer96}, scale  with $1/\Omega T$ 
where $\hbar \Omega$ is a characteristic interaction energy and $T$ is the 
effective time required for the transfer. In some potential applications, as 
for example the transfer of information in form of coherences \cite 
{Pellizzari95}, it is desirable to minimize these losses without the need 
of intense pulses or long transfer times. Intense fields induce time-varying 
ac-Stark shifts, which may be detrimental to the coherence transfer.  
Short times are required to minimize the effect of 
decoherence processes during the transfer \cite{cavity_decay}. 
 
An approach, which reduces non-adiabatic losses for pulses of moderate 
fluence in a three-state system, was recently introduced in Ref.\cite{loop}. 
In addition to the pair of Raman pulses (``pump pulse'' and ``Stokes 
pulse'') which couple the initial and target state via a common upper level, 
a direct coupling (called ``detuning pulse'') between them is introduced. 
This scheme of loop-STIRAP does not require the usual adiabaticity 
conditions (of large pulse areas), nor is it of the $\pi$-pulse type 
(requiring specific pulse areas). Nevertheless, the scheme can produce 
complete population transfer. 
 
In the present paper we show that the physical mechanism of loop-STIRAP is 
not an adiabatic rotation of the dark state, but the rotation of a {\it % 
higher-order trapping state} in a generalized adiabatic basis. The concept 
of generalized adiabatic basis sets allows to rationalize many 
examples of population transfer even when the adiabaticity condition is 
poorly fulfilled. If pump and Stokes pulses fulfill certain conditions (they 
are then called generalized matched pulses), a higher-order trapping exists, 
which is an exact constant of motion. In this case analytic solutions for 
the atomic dynamics can be found which in contrast to the case of ordinary 
matched pulses with identical pulse shape \cite{matched} also include the 
possibility of population transfer. This can be exploited to design pulse 
sequences which give maximum population transfer. In contrast to techniques 
based on optimum control theory, which are used for such tasks, the 
generalized-dark-state concept provides a physical interpretation of the 
results. However, the design of pulse, which in some cases can lead to 
complete 
population transfer (i.e. without {\it any} diabatic losses) needs to 
 respect more restrictive 
requirements for specific pulse properties similar to $\pi$-pulse techniques. 
 
Our paper is organized as follows. In Sec.II we discuss the loop-STIRAP and 
propose a simple physical interpretation  in terms of an adiabatic rotation 
of a generalized trapping state. In Sec.III we define  generalized trapping 
states via an iterative partial diagonalization of the time-dependent 
Hamiltonian. In Sec.IV we derive conditions under which a higher-order 
trapping state is an exact constant of motion and thus allow for an 
analytic solution of the atomic dynamics. Finally, various examples of 
population and 
coherence transfer based on generalized trapping states are discussed 
in Sec.V. 
 
%%%%%%%%%%%%%%%%%%%%%%%%%%%%%%%%%%%%%%%%%%%%%%%%%%%%%%%%%%%%%%%%%%%%%%%%%%%%%% 
 
\section{Loop-STIRAP} 
 
%%%%%%%%%%%%%%%%%%%%%%%%%%%%%%%%%%%%%%%%%%%%%%%%%%%%%%%%%%%%%%%%%%%%%%%%%%%%%% 
 
\begin{figure}[tbp]
\begin{center}
\leavevmode \epsfxsize=6 true cm
\epsffile{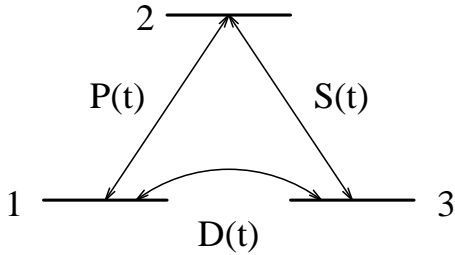}
\end{center}
\caption{Three-state system with loop linkage. $P(t)$, $S(t)$, $D(t)$ denote 
Rabi-frequencies of pump, Stokes and detuning pulse.} 
\label{loop_system} 
\end{figure}

To set the stage we consider in the present section a three-state system 
driven by coherent fields in a loop configuration, as shown in Fig.~\ref 
{loop_system}. The bare atomic states $\psi_1$ and $\psi_3$ are coupled by a 
resonant Raman transition via the excited atomic state $\psi_2$ by a pump 
pulse and a Stokes pulse, having Rabi-frequencies $P(t)$ 
and $S(t)$, respectively, which are in general complex. In addition there 
is a direct coupling between 
states $1$ and $3$ by a coherent detuning pulse described by the (complex) 
Rabi-frequency $D(t)$. Before the application of the pulses the system is in 
state $1$ and the goal is to transfer all population  into the target state $% 
3$ by an appropriate sequence of pulses. For simplicity we assume that the 
carrier frequencies of the pulses coincide with the atomic transition 
frequencies and that the phases of the pulses are time-independent. Since 
the phases of pump and Stokes fields can be included into the definition of 
the bare atomic states $\psi_1$ and $\psi_3$, they can be set equal to zero 
without loss of generality. The phase of the detuning pulse is relevant and 
cannot be 
eliminated. The time-dependent Schr\"odinger equation for 
this system, in the usual rotating wave approximation, reads  
\begin{equation} 
\frac{d}{dt}{\bf C}(t) = -i\, {\sf W}(t) {\bf C}(t) 
\end{equation} 
where ${\bf C}(t)$ is the column vector of probability amplitudes $% 
C_n(t)=\langle n |\psi(t)\rangle$, ($|n\rangle\in\{\psi_1, \psi_2, \psi_3\}$% 
). The evolution matrix ${\sf W}(t)$ has the form  
\begin{equation} 
{\sf W}(t)=\frac{1}{2} \left[ \matrix{0& P(t) & D(t)\cr P(t)&0&S(t)\cr 
D^*(t)&S(t)&0} \right] .  \label{W_loop} 
\end{equation} 
 
%%%%%%%%%%%%%%%%%%%%%%  figure 1 %%%%%%%%%%%%%%%%%%%%%%%%%%%%%% 
 
%%%%%%%%%%%%%%%%%%%%%%%%%%%%%%%%%%%%%%%%%%%%%%%%%%%%%%%%%%%%%% 
 
It is well known that 
the counterintuitive pulse sequence (Stokes puls precedes pump pulse,
without a detuning pulse) leads to an almost complete 
population transfer, if the adiabaticity condition $\Omega T \gg 1$ is 
fulfilled. Here $T$ is the characteristic time for the transfer, given by 
the interval where $S(t)$ and $P(t)$ overlap, and $\Omega$ the 
effective total Rabi-frequency averaged over the interval $T$  
\begin{equation}
\Omega = \frac{1}{T} \int_{-\infty}^{\infty}\!\! dt \sqrt{ P(t)^2 + S(t)^2 }. 
\end{equation}
 
\begin{figure}[tbp] 
\begin{center}
\leavevmode \epsfxsize=7 true cm
\epsffile{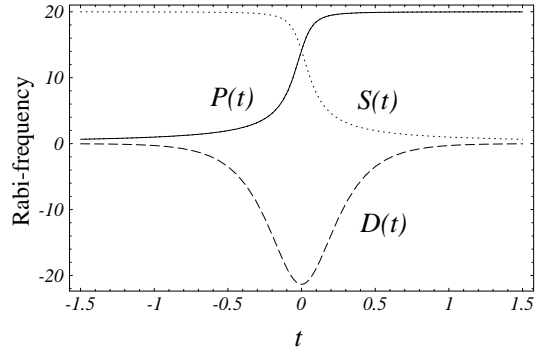}
\end{center}
\caption{Pair of ramped pump (line) and Stokes (dotted) pulses with $% 
A_P=A_S=20$ and $T_P=T_S=0.1$ applied in counterintuitive order (Stokes 
precedes pump) with additional hyperbolic secant detuning pulse (dashed) 
with $A_D=-13.4 i$ and $T_D =0.2$ } 
\label{loop_pulses} 
\end{figure}

As shown in \cite{loop} an almost perfect transfer is also possible when 
pump and Stokes alone do not fulfill the adiabaticity condition by applying 
an additional detuning pulse. Fig. \ref{loop_pulses} illustrates an example 
of ramped pump and Stokes pulses intersected by a hyperbolic-secant detuning 
pulse,  
\begin{eqnarray} 
P(t) &=& A_P\, \sin\Bigl[\frac{1}{2}\arctan\bigl(t/T_P\bigr)+ \frac{\pi}{4}% 
\Bigr], \\ 
S(t) &=& A_S\, \cos\Bigl[\frac{1}{2}\arctan\bigl(t/T_S\bigr)+ \frac{\pi}{4}% 
\Bigr], \\ 
D(t) &=& A_D\, {\rm sech}\bigl[t/T_D\bigr]. 
\end{eqnarray} 
 
%%%%%%%%%%%%%%%%%%%%%%%%%%%%%%%%%%%%%%%%%%%%%%%%%%%%%%%%%%%%%% 

%%%%%%%%%%%%%%%%%%%%%%%%%%%%%%%%%%%%%%%%%%%%%%%%%%%%%%%%%%%%%% 
 
\begin{figure}[tbp] 
\begin{center}
\leavevmode \epsfxsize=8 true cm
 \epsffile{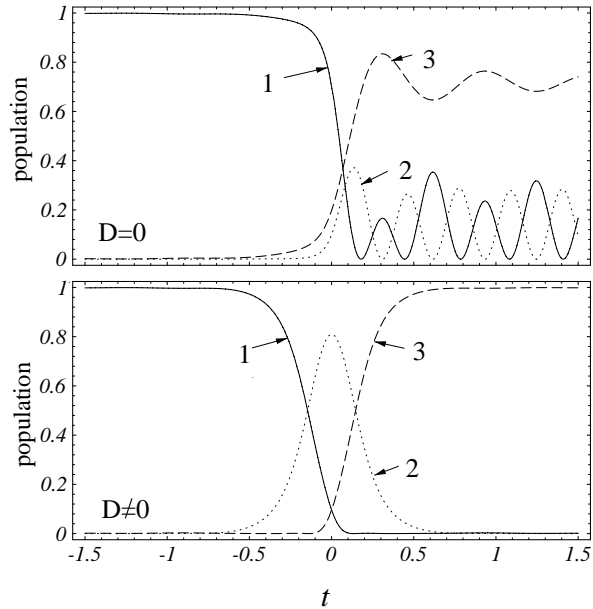}
\end{center}
\caption{Populations of states $\psi_1$ (line), $\psi_2$ (dotted) and $\psi_3 
$ (dashed) for pulse sequence of Fig.\ref{loop_pulses}. The upper picture 
shows population when only pump and Stokes pulses are applied, and the lower 
one if the detuning pulse is added. } 
\label{loop_populations} 
\end{figure}

Fig. \ref{loop_populations} shows examples of population histories for these 
pulses. When only the pump and Stokes pulses are present, the population 
transfer is rather poor, since the pulse areas are small ($\Omega\, T\sim 
|A_P|\, T_P = |A_S|\, T_S =2$). As can be seen from the upper part of Fig.% 
\ref{loop_populations}, only about 70\% of the initial population ends up in 
state $3$. 
 
%%%%%%%%%%%%%%%%%%%%%%%%%%%%%%%%%%%%%%%%%%%%%%%%%%%%%%%%%%%%%% 

%%%%%%%%%%%%%%%%%%%%%%%%%%%%%%%%%%%%%%%%%%%%%%%%%%%%%%%%%%%%%% 
 
The situation is remarkably different when a detuning pulse with $% 
|A_D|T_D\approx 2.7$ and a phase factor of ${\rm e}^{-i\pi/2}$ is applied; see the 
lower part of Fig.\ref{loop_populations}. With a detuning pulse present all 
the population is transfered from the initial to the target state. This 
result is relatively insensitive to changes in the amplitude (or the shape 
of the detuning pulse) if the phase is $-\pi/2$.

We note that in contrast to ordinary STIRAP there is (for a short time) a 
substantial intermediate population of state 2. This indicates that the 
transfer does not occur as adiabatic rotation of the dark state from $\psi_1$ 
to $\psi_3$. 
 
For our present discussion it is useful to describe ordinary STIRAP in terms 
of the following set of adiabatic superposition states  
\begin{equation} 
\left[ \matrix{\ket{\Phi_1(t)}\cr\ket{\Phi_2(t)}\cr\ket{\Phi_3(t)}} \right] =% 
{\sf U}(t)^* \left[ \matrix{\ket{\psi_1}\cr\ket{\psi_2}\cr\ket{\psi_3}} 
\right]  \label{dark} 
\end{equation} 
with the unitary matrix  
\begin{equation} 
{\sf U}(t)= \left[ \matrix{0 & 1 & 0 \cr \enspace \sin\theta_0(t) & 0 & 
\quad \cos\theta_0(t) \cr i\cos\theta_0(t) & 0 & -i\sin\theta_0(t) } \right] 
. 
\end{equation} 
The dynamical angle $\theta_0$ is defined by  
\begin{eqnarray} 
\tan\theta_0(t) = \frac{P(t)}{S(t)}.  \label{theta0} 
\end{eqnarray} 
The vector of probability amplitudes in the bare atomic basis ${\bf C}(t)$ 
and a corresponding vector ${\bf B}(t)$ in the superposition basis (\ref 
{dark}) are related through the transformation  
\begin{equation} 
{\bf B}(t) ={\sf U}(t){\bf C}(t). 
\end{equation} 
 
Since ${\sf U}(t)$ is time-dependent, the transformed evolution matrix has 
the form  
\begin{equation} 
{\sf W}(t)\to \widetilde {{\sf W}}(t)={\sf U}(t){\sf W}(t){\sf U}(t)^{-1} 
+i\, \dot {{\sf U}}(t){\sf U}(t)^{-1}. 
\end{equation} 
In the adiabatic limit, the second term can be disregarded and we are left 
with the first one, which for ordinary STIRAP, i.e. without the detuning 
pulse, reads  
\begin{equation} 
{\sf U}(t){\sf W}(t){\sf U}(t)^{-1} =\frac{1}{2} \left[ \matrix{0 & 
\Omega(t) & 0 \cr \Omega(t) & 0 & 0\cr 0 & 0 & 0} \right] , 
\end{equation} 
where $\Omega(t)=\sqrt{P(t)^2+S(t)^2}$. One recognizes that the superposition 
state $\Phi_3(t)$ is decoupled from the coherent interaction in this limit. 
Moreover, because $\Phi_3(t)$ does not contain the excited atomic state $% 
\psi_2$, it does not spontaneously radiate and is therefore called a dark 
state \cite{Arimondo}. For a counterintuitive sequence of pulses the angle $% 
\theta_0(t)$ vanishes initially and approaches $\pi/2$ for $t\to\infty$. 
Thus $\Phi_3(t) $ asymptotically coincides with the initial and target 
states for $t\to \pm \infty$ respectively. Therefore ordinary STIRAP can be 
understood as a rotation of the adiabatic dark state $\Phi_3(t)$ from the 
initial to the target bare atomic state \cite{STIRAP2}. 
Non-adiabatic corrections are 
contained in the second contribution to $\widetilde {{\sf W}}(t)$  
\begin{equation} 
i\, \dot{{\sf U}}(t){\sf U}(t)^{-1} = \frac{1}{2} \left[ \matrix{0 & 0 & 
0\cr 0 & 0 & 2\dot\theta_0(t) \cr 0 & 2\dot\theta_0(t) & 0} \right] . 
\end{equation} 
They give rise to a coupling between the dark state $\Phi_3(t)$ and the 
so-called bright state $\Phi_2(t)$. 
 
Let us now apply the same transformation to the loop-STIRAP system, i.e. 
including the detuning pulse. We find:  
\begin{equation} 
\widetilde{{\sf W}}(t) =\frac{1}{2} \left[ \matrix{0 & \Omega(t) & 0 \cr 
\Omega(t) & {\rm Re}\bigl[D(t)\bigr]\sin 2\theta_0(t) & 2{\dot\theta}_0(t) 
+i\bigl[D(t)\sin^2\theta_0(t) - D^*(t)\cos^2\theta_0(t)\bigr] \cr 0 & 
2{\dot\theta}_0(t) -i\bigl[D^*(t)\sin^2\theta_0(t) - 
D(t)\cos^2\theta_0(t)\bigr] & -{\rm Re}\bigl[D(t)\bigr]\sin 2\theta_0(t)} 
\right] . 
\end{equation} 
If $D(t)$ is real or complex but not strictly imaginary, there is a time dependent energy shift of the superposition 
states $\Phi_2(t)$ and $\Phi_3(t)$ and the detuning pulse adds an imaginary 
part to the nonadiabatic coupling. If $D(t)$ is imaginary, as in the example 
discussed above, there is no detuning but a {\it real} contribution to the 
nonadibatic coupling. Let us now assume an imaginary detuning pulse, i.e. $% 
D(t) = i \widetilde{D}(t)$, with $\widetilde{D}(t)$ being real. In this case 
the transformed evolution matrix simplifies to  
\begin{equation} 
{\widetilde{{\sf W}}(t)} = \frac{1}{2} \left[ \matrix{0 & \Omega(t) & 0 \cr 
\Omega(t) & 0 & 2 \dot\theta_0(t)-\widetilde{D}(t)\cr 0 & 
2\dot\theta_0(t)-\widetilde{D}(t) & 0} \right] .  \label{W_dressed} 
\end{equation} 
 
If the amplitude of the detuning pulse matches the non-adiabatic term, i.e. 
if $\widetilde{D}(t)=2\dot\theta_0(t)$, the dark state $\Phi_3$ is exactly 
decoupled even if the adiabaticity condition for pump and Stokes alone ($% 
\Omega(t)$ being much larger than $\dot\theta_0(t)$) is not fulfilled. 
However, since $\theta_0(t)$ rotates from $0$ to $\pi/2$, the detuning pulse 
would have to be exactly a $\pi$-pulse in such a case.  
\begin{equation} 
\int_{-\infty}^\infty\!\! dt\, {\widetilde D}(t)=
\int_{-\infty}^\infty\!\! dt\, 2 {\dot\theta}_0(t) = 2 
\theta_0(t)\Bigr\vert_{-\infty}^{+\infty} =\pi 
\end{equation} 
Furthermore no pump or Stokes pulses were required for population transfer 
to begin with, since at any time the entire population is kept in the dark 
state by the action of the detuning pulse and thus pump and Stokes would not 
interact with the atoms. This is consistent with the observation that an 
exactly decoupled state $\Phi_3$ implies exactly vanishing (not only 
adiabatically small!) probability amplitude of the excited bare state $\psi_2 
$ for all times. Since the origin of population transfer in this case is the 
well-known phenomenon of $\pi$-pulse coupling, which requires a careful 
control of the area and the shape of the detuning pulse, the case $% 
\widetilde{D}(t)=2\dot\theta_0(t)$ is of no further interest here. 
 
\begin{figure}[tbp] 
\begin{center}
\leavevmode \epsfxsize=7 true cm
\epsffile{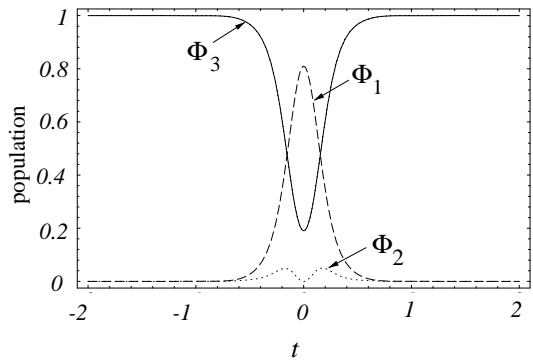}
\end{center}
\caption{Population of superposition states $\Phi_1$ (dashed), $\Phi_2$ 
(dotted), and the dark state $\Phi_3$ (line). Parameters are that of Fig.\ref 
{loop_pulses} } 
\label{loop_dressed_pop} 
\end{figure}

On the other hand, if $\widetilde{D}(t)$ is negative, as in the example of 
Fig.\ref{loop_pulses}, the non-adiabatic coupling is effectively increased 
by the detuning pulse (note that $d\theta_0(t)/dt >0$). Thus the success of 
population transfer in Fig.\ref{loop_populations} cannot be understood as 
dark-state rotation. This is illustrated in Fig.\ref{loop_dressed_pop}, 
which shows the populations of the superposition states $\Phi_1=\psi_2$, $% 
\Phi_2$, and $\Phi_3$ for the above example. One clearly sees that about 
80\% of the population is driven out of the dark state during the 
interaction. 
 
%%%%%%%%%%%%%%%%%%%%%%%%%%%%%%%%%%%%%%%%%%%%%%%%%%%%%%%%%%%%%% 

%%%%%%%%%%%%%%%%%%%%%%%%%%%%%%%%%%%%%%%%%%%%%%%%%%%%%%%%%%%%%% 
 
It is worth noting, however, that $\Phi_2$ remains almost unpopulated during 
the interaction and all population exchange happens between states $\Phi_1$ 
and $\Phi_3$. This suggests an interpretation of the process as {\it % 
adiabatic population return between the superposition states} $\Phi_1$ and $% 
\Phi_3$. In fact comparing the dressed-state evolution matrix $\widetilde{% 
{\sf W}}(t)$, Eq.(\ref{W_dressed}), with the bare-state evolution matrix $% 
{\sf W}(t)$, Eq.(\ref{W_loop}) (without detuning pulse), one recognizes a 
formal agreement with the correspondence $P(t) \leftrightarrow \Omega(t)$ 
and $S(t) \leftrightarrow 2\dot\theta_0(t)-\widetilde {D}(t)$. That is there 
exists a {\it generalized trapping state} which is a superposition of the 
states $\Phi_1$ and $\Phi_3$. Since here $\Omega(t)=$ const. 
and $2\dot{\theta}_0(t)-\widetilde{D}(t)$ vanishes in the asymptotic limits $% 
t\to\pm\infty$, this generalized trapping state coincides with $\Phi_3$ for $% 
t\to\pm\infty$, which in turn coincides with $\psi_1$ and $\psi_3$ in the 
respective limits. 
 
To quantify this statement let us introduce a basis of {\it second-order 
adiabatic states}. Using now the first-order states $\Phi_1$, $\Phi_2$, and $% 
\Phi_3$ as a basis set instead of the bare atomic states, we introduce in 
analogy to Eq.(\ref{dark})  
\begin{equation} 
\left[ \matrix{\ket{\Phi_1^{(2)}(t)}\cr\ket{\Phi_2^{(2)}(t)}\cr 
\ket{\Phi_3^{(2)}(t)}} \right] ={\sf U}_1(t)^* \left[ \matrix{\ket{% 
\Phi_1(t)}\cr\ket{\Phi_2(t)}\cr \ket{\Phi_3(t)}} \right] = {\sf U}% 
_1(t)^*\cdot{\sf U}(t)^* \left[ \matrix{\ket{\psi_1}\cr\ket{\psi_2}\cr 
\ket{\psi_3}} \right] . 
\end{equation} 
The unitary transformation matrix is given by  
\begin{eqnarray} 
{\sf U}_1(t)= \left[ \matrix{0 & 1 & 0 \cr \enspace\sin\theta_1(t) & 0 & 
\quad\cos\theta_1(t) \cr i\cos\theta_1(t) & 0 & -i\sin\theta_1(t) } \right] . 
\end{eqnarray} 
which has the same form as ${\sf U}(t)$, Eq.(\ref{W_loop}) 
but here the dynamical angle $\theta_1(t)$ is defined by  
\begin{eqnarray} 
\tan\theta_1(t) = \frac{\Omega(t)}{2\dot\theta_0(t)-\widetilde{D}(t)}. 
\end{eqnarray} 
Denoting the vector of probability amplitudes in these generalized adiabatic 
states by ${\bf B}^{(2)}(t)$ we find the relation  
\begin{equation} 
{\bf B}^{(2)}(t) ={\sf U}_1(t) {\bf B}(t). 
\end{equation} 
 
One easily verifies that for the above example more than 95\% of the 
population remains in the generalized trapping state $\Phi_3^{(2)}(t) 
$. Thus the success of the population transfer in loop STIRAP can be 
understood as a rotation of the second-order decoupled state $\Phi_3^{(2)}(t) 
$ -- which is an approximate constant of motion -- from the initial to the 
target bare atomic state. 
 
%%%%%%%%%%%%%%%%%%%%%%%%%%%%%%%%%%%%%%%%%%%%%%%%%%%%%%%%%%%%%%%%%%%%%%%%%%% 
 
\section{generalized adiabatic basis and generalized trapping states for 
STIRAP} 
 
%%%%%%%%%%%%%%%%%%%%%%%%%%%%%%%%%%%%%%%%%%%%%%%%%%%%%%%%%%%%%%%%%%%%%%%%%% 
 
\begin{figure}[tbp] 
\begin{center}
\leavevmode \epsfxsize=7 true cm
\epsffile{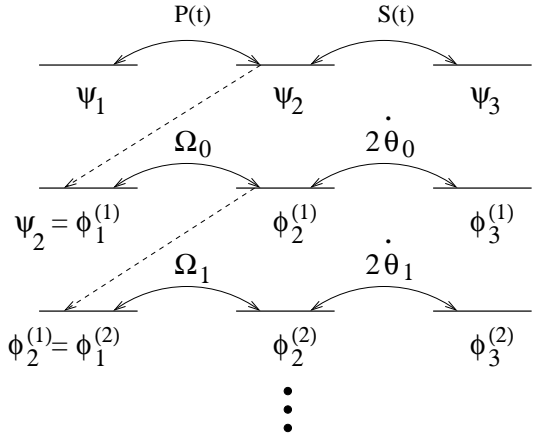}
\end{center}
\caption{Iterative definition of $n$th order adiabatic basis } 
\label{iteration} 
\end{figure}

We now return to the case of ordinary STIRAP, i.e. without a detuning pulse $% 
D$. The formal equivalence of ${\sf W}(t)$ and $\widetilde{{\sf W}}(t)$ 
suggest an iteration of the procedure introduced in the last section. We 
define an $n$th order generalized adiabatic basis by the iteration:  
\begin{equation} 
\left[ \matrix{\ket{\Phi_1^{(n)}(t)}\cr\ket{\Phi_2^{(n)}(t)}\cr\ket{% 
\Phi_3^{(n)}(t )}} \right] = {\sf U}_{n-1}(t)^* \left[ \matrix{\ket{% 
\Phi_1^{(n-1)}(t)}\cr \ket{\Phi_2^{(n-1)(t)}}\cr \ket{\Phi_3^{(n-1)}(t)}} 
\right] ={\sf U}_{n-1}(t)^* \cdot{\sf U}_{n-2}(t)^*\cdots{\sf U}_0(t)^* 
\left[ \matrix{\ket{\psi_1}\cr\ket{\psi_2}\cr\ket{\psi_3}} \right] . 
\end{equation} 
Correspondingly we obtain for the vector of probability amplitudes in the $n$% 
th order basis  
\begin{equation} 
{\bf B}^{(n)} = {\sf U}_{n-1} {\bf B}^{(n-1)} = {\sf U}_{n-1}\cdot {\sf U}% 
_{n-2} \cdots {\sf U}_0 {\bf C}\equiv {\sf V}_n {\bf C}  \label{B-C} 
\end{equation} 
where we have dropped the time dependence. The $n$th order transformation 
matrix is defined as  
\begin{equation} 
{\sf U}_n(t) \equiv \left[ \matrix{0 & 1 & 0 \cr \enspace\sin\theta_n(t) & 0 
& \quad\cos\theta_n(t) \cr i\cos\theta_n(t) & 0 & -i\sin\theta_n(t) } 
\right] , 
\end{equation} 
with  
\begin{eqnarray} 
\sin\theta_0(t)&=&\frac{P(t)}{\Omega_0(t)},\qquad\quad\cos\theta_0(t)\;=  
\frac{S(t)}{\Omega_0(t)},\qquad\quad\enspace \Omega_0(t)\; =\sqrt{% 
P(t)^2+S(t)^2}, \\ 
\sin\theta_n(t)&=&\frac{\Omega_{n-1}(t)}{\Omega_n(t)},\qquad\cos\theta_n(t)=  
\frac{2{\dot\theta}_{n-1}(t)}{\Omega_n(t)},\qquad \Omega_n(t) =\sqrt{% 
\Omega_{n-1}(t)^2+4{\dot\theta}_{n-1}(t)^2}.  \label{thetan} 
\end{eqnarray} 
The iteration is illustrated in Fig.\ref{iteration}. 
 
%%%%%%%%%%%%%%%%%%%%%%%%%%%%%%%%%%%%%%%%%%%%%%%%%%%%%%%%%%%%%%%%%%%%%%%%%%% 
 
%%%%%%%%%%%%%%%%%%%%%%%%%%%%%%%%%%%%%%%%%%%%%%%%%%%%%%%%%%%%%%%%%%%%%%%%%%% 
 
In the $n$th-order basis, the equation of motion has then the form  
\begin{equation} 
\frac{d}{dt}{\bf B}^{(n)}(t) =-i{\sf W}_n(t) {\bf B}^{(n)}(t), 
\end{equation} 
with  
\begin{eqnarray} 
{\sf W}_n(t) &\equiv& \frac{1}{2} \left[ \matrix{ 0 & \Omega_n(t) 
\sin\theta_n(t) & 0 \cr \Omega_n(t) \sin\theta_n(t) & 0 & 
\Omega_n(t)\cos\theta_n(t) \cr 0 & \Omega_n(t)\cos\theta_n(t) & 0 } \right] 
\\ 
&=& \frac{1}{2} \left[ \matrix{ 0 & \Omega_{n-1}(t) & 0 \cr \Omega_{n-1}(t) 
& 0 & 2{\dot\theta}_{n-1}(t) \cr 0 & 2{\dot\theta}_{n-1}(t) & 0 } \right] .  
\nonumber 
\end{eqnarray} 
If $\cos\theta_k(t)$ vanishes, which  implies that $\theta_{k-1}$ is 
time-independent, the state $\Phi_3^{(k)}$ decouples from the interaction. 
In this case exact analytic solutions of the atomic dynamics can be found as 
discussed in the next section. The analytic solutions also include cases of 
population or coherence transfer. If $\cos\theta_k(t)$ does not vanish but 
is small, the corresponding coupling in the evolution matrix can be treated 
perturbatively. In such a situation we have a {\it generalized adiabatic 
dynamics}. 
 
In conclusion of this section it should be noted, that the iterative 
definition of a generalized adiabatic basis is conceptually very similar to 
the superadiabatic approach of Berry \cite{super} introduced for two-level 
systems. 
 
%%%%%%%%%%%%%%%%%%%%%%%%%%%%%%%%%%%%%%%%%%%%%%%%%%%%%%%%%%%%%%%%%%%%%%%%%%%% 
 
\section{Generalized matched pulses and analytic solution of atomic dynamics} 
 
%%%%%%%%%%%%%%%%%%%%%%%%%%%%%%%%%%%%%%%%%%%%%%%%%%%%%%%%%%%%%%%%%%%%%%%%%%% 
 
If a dynamical angle $\theta_{n-1}$ is a constant, the time-dependent state $% 
\Phi_3^{(n)}(t)$ is decoupled from the interaction (constant of motion). In 
this case the dynamical problem reduces to that of a two-state system 
interacting via a {\it real} resonant coherent coupling plus a decoupled 
state.  
\begin{equation} 
\frac{d}{dt} \left[ \matrix{B_1^{(n)}(t)\cr B_2^{(n)}(t)\cr B_3^{(n)}(t)} 
\right] =-\frac{i}{2} \left[ \matrix{0 & \Omega_{n-1}(t) &0 \cr 
\Omega_{n-1}(t) & 0 & 0 \cr 0 & 0 & 0} \right] \left[ \matrix{B_1^{(n)}(t)% 
\cr B_2^{(n)}(t)\cr B_3^{(n)}(t)} \right] 
\end{equation} 
This equation can immediately be solved  
\begin{eqnarray} 
B_1^{(n)}(t) &=& B_1^{(n)}(0)\, \cos\phi(t) - i B_2^{(n)}(0)\, \sin\phi(t), 
\label{match_sol1} \\ 
B_2^{(n)}(t) &=& B_2^{(n)}(0)\, \cos\phi(t) - i B_1^{(n)}(0)\, \sin\phi(t), 
\label{match_sol2} \\ 
B_3^{(n)}(t) &=& B_3^{(n)}(0),  \label{match_sol3} 
\end{eqnarray} 
where  
\begin{equation} 
\phi(t) = \frac{1}{2}\int_0^t\!\! d\tau\, \Omega_{n-1}(\tau). 
\end{equation} 
In particular if the atom is initially in the trapping state, it will stay in 
that state. 
 
For example if $\theta_0$ does not depend on time, the usual dark state $% 
\Phi_3^{(1)}$ is an exact constant of motion. As can be seen from Eq.(\ref 
{theta0}), for $\theta_0$ to be time-independent, Stokes and pump need to be 
either cw fields or need to have the same envelope function, i.e. have to be  
{\it matched pulses} \cite{matched},  
\begin{eqnarray} 
S(t) &=& \Omega_0(t) \,\cos\theta_0, \\ 
P(t) &=& \Omega_0(t) \, \sin\theta_0, 
\end{eqnarray} 
where $\Omega_0(t)$ can be an arbitrary function of time and 
$\theta_0=$const. The 
atomic dynamics is trivial in this case. Since $% 
\Phi_3^{(1)}$ is time-independent, the trapping state is a constant 
superposition of the bare atomic states $1$ and $3$. 
 
On the other hand, if some higher-order dynamical angle $\theta_n$ is 
constant, the system remains in a generalized trapping state if initially 
prepared in it. The projection of this state onto the bare atomic basis is 
in general time-dependent, and one can have a substantial rearrangement of atomic level population including 
population transfer. If a higher-order dynamical angle is constant we will 
call pump and Stokes pulses {\it generalized matched pulses}. 
 
To obtain an explicit condition for generalized matched pulses in terms of $% 
P(t)$ and $S(t)$  we successively integrate relations (\ref{thetan}). This 
leads to the iteration  
\begin{eqnarray} 
\theta_{k-1}(t) &=& \frac{1}{2}\int_{-\infty}^t\!\!\! dt^\prime\, 
\Omega_k(t^\prime)\, \cos\theta_k(t^\prime)+ \theta_k^0,\cr \Omega_{k-1}(t) 
&=& \Omega_k(t)\, \sin\theta_k(t),  \label{iterate} 
\end{eqnarray} 
starting with some $\theta_n(t)=\theta_n =$ const.~and $\Omega_n(t)$ as  an 
arbitrary function of time. Each iteration leads to one constant $\theta_k^0$% 
, which can be freely chosen. The application of generalized matched pulses 
to coherent population transfer will be discussed in the next section. 
 
As noted before there may be cases, where for some number $n$ the dynamical 
angle $\theta_n(t)$ does depend on time but its time-derivative is much 
smaller than the corresponding generalized Rabi-frequency $\Omega_n(t)$, 
while the same is not true for all $k< n$. In this case the state $% 
\Phi_3^{(n)}(t)$ is an approximate constant of motion and we have an $n$th 
order adiabatic process. The example of loop-STIRAP discussed in the last 
section is a realization of a higher-order adiabatic process, which is 
non-adiabatic in the first-order basis. 
 
%%%%%%%%%%%%%%%%%%%%%%%%%%%%%%%%%%%%%%%%%%%%%%%%%%%%%%%%%%%%%%%%%%%%%%%%%%% 
 
\section{Application of generalized matched pulses to population- and 
coherence transfer} 
 
%%%%%%%%%%%%%%%%%%%%%%%%%%%%%%%%%%%%%%%%%%%%%%%%%%%%%%%%%%%%%%%%%%%%%%%%%%% 
 
In the following we discuss several examples for a coherent transfer of 
population from one non-decaying 
state to the other or to the excited state using 
generalized matched pulses. We furthermore discuss the possibility to 
transfer coherence, for example from the ground state transition to an 
optical transition. Since in all cases there exist a generalized 
trapping state 
which is an exact constant of motion, we can obtain exact analytic results 
for the atomic dynamics. 
 
%------------------------------------------------------------------------- 
 
\subsection{Population and coherence transfer with second-order generalized 
matched pulses} 
 
%------------------------------------------------------------------------- 
 
\subsubsection{Complete transfer of coherence from a ground-state doublet to 
an optical transition} 
 
%------------------------------------------------------------------------- 
 
First we discuss the case when $\Phi_3^{(2)}$ is an exact constant 
of motion, i.e. a trapping state. Furthermore we assume that the state 
vector $\Psi$ coincides with this trapping state at $t=-\infty$. Then the 
system will remain in the trapping state at later times. Therefore $\theta_1=$const. and it is 
clear from Fig.\ref{iteration} that $\Psi$ is a time independent 
superposition of states $\Phi^{(1)}_1$ and $\Phi^{(1)}_3$ and thus has at 
all times a constant probability amplitude of the bare atomic state 2. In 
fact from  
\begin{equation} 
{\bf C}={\sf V}_2^{-1} {\bf B}^{(2)} = {\sf V}_2^{-1} \left[ \matrix{0\cr 
0\cr 1} \right] 
\end{equation} 
we find  
\begin{eqnarray} 
\left[ \matrix{C_1(t)\cr C_2(t)\cr C_3(t)} \right] =-i\cos\theta_1 \left[ % 
\matrix{\enspace i\tan\theta_1\, \cos\theta_0(t) \cr 1\cr -i \tan\theta_1\, 
\sin\theta_0(t) } \right] . 
\end{eqnarray} 
We now identify state 2 with a lower i.e. non-decaying level and state 3 
with an excited states. The pump pulse $P(t)$ then couples two ground states 
which could be realized for example by a magnetic coupling. The Stokes 
pulse, which couples states 2 and 3 is considered an optical pulse. Due to 
the finite and constant admixture of state 2 to the trapping state, 
second-order generalized matched pulses are best suited to transfer 
coherence for example from the 1-2 transition to the 3-2 transition. 
 
We now want to construct pulses, that would lead to the desired complete 
coherence transfer. To achieve this we have to satisfy the initial and final 
conditions  
\begin{eqnarray} 
\theta _{0}(-\infty ) &=&0,  \label{cond11} \\ 
\theta _{0}(+\infty ) &=&\pi /2.  \label{cond12} 
\end{eqnarray} 
On the other hand, the iteration equation (\ref{iterate}) requires for 
second-order matched pulses that 
\begin{eqnarray} 
\theta _{0}(t) &=&\frac{1}{2}\int_{-\infty }^{t}\!\!\!dt^{\prime }\,\Omega 
_{1}(t^{\prime })\cos \theta _{1}+\theta _{0}^{0},  \label{theta0_cond} \\ 
\Omega _{0}(t) &=&\Omega _{1}(t)\sin \theta _{1}, 
\end{eqnarray} 
where $\theta _{1}$ and $\theta _{0}^{0}$ are arbitrary constants and $% 
\Omega _{1}(t)$ an arbitrary positive function of time. To fulfill the 
initial condition (\ref{cond11}) we set $\theta _{0}^{0}=0$. In order to 
satisfy the final condition (\ref{cond12}) we then have to adjust the total 
pulse area (see Eq.(\ref{theta0_cond}))  
\begin{equation} 
A_{0}=\int_{-\infty }^{\infty }\!\!\!dt\,\Omega _{0}(t)=\pi \tan \theta _{1}. 
\label{area1} 
\end{equation} 
Thus pump and Stokes pulses have the form  
\begin{eqnarray} 
P(t) &=&\Omega _{0}(t)\sin \biggl[\frac{\pi A\left( t\right) }{2A_{0}}% 
\biggr], \\ 
S(t) &=&\Omega _{0}(t)\cos \biggl[\frac{\pi A\left( t\right) }{2A_{0}}% 
\biggr]. 
\end{eqnarray} 
with 
\begin{equation} 
A\left( t\right) =\int_{-\infty }^{t}\!\!\!dt^{\prime }\,\Omega 
_{0}(t^{\prime }) 
\end{equation} 
With this choice an initial coherent superposition of states 2 and 1  
\begin{equation} 
\Psi (-\infty )=-i\cos \theta _{1}\,\psi _{2}+\sin \theta _{1}\,\psi _{1} 
\label{initial11} 
\end{equation} 
can be completely mapped into a coherent superposition of states 2 and 3  
\begin{equation} 
\Psi (+\infty )=-i\cos \theta _{1}\,\psi _{2}-\sin \theta _{1}\,\psi _{3}. 
\label{final11} 
\end{equation} 
In order to transfer a given ground-state coherence to an optical transition 
the pulse area $A_{0}$ should be chosen according to (\ref{area1}), $% 
A_{0}=|C_{1}(-\infty )/C_{2}(-\infty )|$. The   shape of $\Omega (t)$ is 
otherwise arbitrary. It should be noted that Eq.(\ref{initial11}) requires a 
certain fixed phase of the initial coherent superposition. The phase of the 
pump pulse, which is included in the definition of $\psi _{1}$ (cf. Sec.II), 
may need adjustment to satisfy this condition.  
 
\begin{figure}[tbp] 
\begin{center}
\leavevmode \epsfxsize=7 true cm
\epsffile{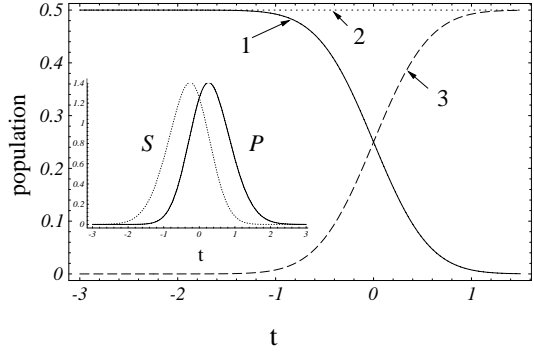}
\end{center}
\caption{Example of complete coherence transfer from $1-2$ to $3-2$ with 
second-order generalized matched pulses. Plotted are the populations of bare 
atomic states for a pair of pulses as shown in the insert. One recognizes 
constant population in state 2 and complete transfer of population in 1 to 3. 
} 
\label{second_order_matched} 
\end{figure}

In Fig.\ref{second_order_matched} we have shown the populations of the bare 
atomic states for the example $\Omega_0(t) =\sqrt{\pi} \exp(-t^2)$ ($A=\pi$) 
and $\Psi(-\infty)=1/\sqrt{2} \bigl(\psi_1 -i \psi_2\bigr)$ from a numerical 
solution of the Schr\"odinger equation. One clearly sees that all population 
from state 1 is transferred to state 3. This transfer happens without 
diabatic losses despite the fact that $A=\pi$ and thus the usual 
adiabaticity condition is only poorly fulfilled. 
 
%%%%%%%%%%%%%%%%%%%%%%%%%%%%%%%%%%%%%%%%%%%%%%%%%%%%%%% 
 
%%%%%%%%%%%%%%%%%%%%%%%%%%%%%%%%%%%%%%%%%%%%%%%%%%%%%%%%% 
 
The process discussed here may have some interesting applications, since it 
allows to transfer coherence from a robust and long-lived ground state 
transition to an optically accessible transition. 
 
The population transfer from 1 to 3 with finite constant state amplitude in 
2 discussed here coincides with the solution found by Malinovsky and Tannor  
\cite{Malinovsky} with numerical optimization techniques. Assuming a finite 
constant amplitude in state 2, these authors numerically optimized the peak 
Rabi-frequency (which in this case is the only remaining free parameter) to 
achieve maximum population transfer. They found, that in order to maximize 
the final amount of population in state 3, the peak Rabi-frequency has to be 
larger than a certain critical value. This can very easily be verified from 
the generalized matched-pulse solutions (\ref{area1},\ref{final11}).  
\begin{eqnarray} 
|C_3(\infty)|^2 &=& \sin^2\theta_1=\frac{A^2}{\pi^2+A^2} \\ 
|C_2(\infty)|^2 &=& \cos^2\theta_1=\frac{\pi^2}{\pi^2+A^2}. 
\end{eqnarray} 
In the limit $\theta_1\to\pi/2$, which implies $A\to\infty$, the admixture 
of level 2 vanishes and we essentially have population transfer from state 1 
to state 3. 
 
%---------------------------------------------------------------------- 
 
\subsubsection{Population transfer from 1 to 3 and non-exponential diabatic 
losses} 
 
%---------------------------------------------------------------------- 
 
\begin{figure}[tbp] 
\begin{center}
\leavevmode \epsfxsize=7 true cm
\epsffile{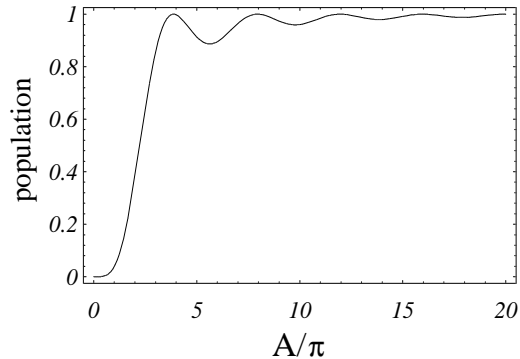}
\end{center}
\caption{Final population in state 3 as a function of total pulse
area $A/\pi$ for population transfer from state 1 with second-order
matched pulses. For $A/\pi=\sqrt{16 n^2-1}$ the transfer is complete
(100.00\%)}
\label{second_order_success}
\end{figure}

We have seen in the last subsection that second-order matched pulses can be 
used to effectively transfer population from state 1 to 3, if there is an 
initial admixture of the excited state. This amplitude is inversely 
proportional to the square of the pulse area $A$. Therefore one could expect 
a good transfer for large $A$ also if all population is initially in state 
1. In this case there is some finite amount of population which is not 
trapped in the generalized dark state $\Phi _{3}^{(2)}$. Clearly in order to 
achieve maximum population transfer, pump and Stokes pulse should be in 
counterintuitive order and hence conditions (\ref{cond11}) and (\ref{cond12}% 
) should be fulfilled. Since the pulses are assumed to be second-order 
matched pulses, the dynamical problem with the initial condition  
\begin{equation} 
\left[ \matrix{B_1^{(2)}(-\infty)\cr B_2^{(2)}(-\infty)\cr 
B_3^{(2)}(-\infty)}\right] ={\sf U}_{1}\cdot {\sf U}_{0}\,\left[ % 
\matrix{1\cr 0\cr 0}\right] =\left[ \matrix{0\cr i\cos\theta_1\cr 
\sin\theta_1}\right]  
\end{equation} 
can easily be solved (see Eq.(\ref{match_sol1}-\ref{match_sol3})). From Eqs.(% 
\ref{thetan}) we find $\Omega _{1}(t)=\Omega _{0}(t)/\sin \theta _{1}$. Thus  
\begin{eqnarray} 
B_{1}^{(2)}(\infty ) &=&\frac{\pi }{\sqrt{\pi ^{2}+A^{2}}}\,\sin \left[  
\frac{1}{2}\sqrt{\pi ^{2}+A^{2}}\right] , \\ 
B_{2}^{(2)}(\infty ) &=&i\frac{\pi }{\sqrt{\pi ^{2}+A^{2}}}\,\cos \left[  
\frac{1}{2}\sqrt{\pi ^{2}+A^{2}}\right] , \\ 
B_{3}^{(2)}(\infty ) &=&\frac{A}{\sqrt{\pi ^{2}+A^{2}}}, 
\end{eqnarray} 
where $A$ is the total pulse area defined in (\ref{area1}). From this we 
find the asymptotic populations of the bare atomic states  
\begin{eqnarray} 
\Bigl|C_{1}(\infty )\Bigr|^{2} &=&\frac{\pi ^{2}}{\pi ^{2}+A^{2}}\,\sin 
^{2}\left( \frac{1}{2}\sqrt{\pi ^{2}+A^{2}}\right) , \\ 
\Bigl|C_{2}(\infty )\Bigr|^{2} &=&\frac{4\pi ^{2}A^{2}}{\bigl(\pi ^{2}+A^{2}% 
\bigr)^{2}}\,\sin ^{4}\left( \frac{1}{4}\sqrt{\pi ^{2}+A^{2}}\right) , \\ 
\Bigl|C_{3}(\infty )\Bigr|^{2} &=&\frac{1}{\bigl(\pi ^{2}+A^{2}\bigr)^{2}}% 
\left[ A^{2}+\pi ^{2}\cos \left( \frac{1}{2}\sqrt{\pi ^{2}+A^{2}}\right) 
\right] ^{2}. 
\end{eqnarray} 
Thus the diabatic losses scale in general with $1/A^{2}$, i.e. 
non-exponentially with $A$. Furthermore for  
\begin{equation} 
\frac{1}{2}\sqrt{\pi ^{2}+A^{2}}=2n\pi \qquad {\rm or}\qquad A=\pi \sqrt{% 
16n^{2}-1} 
\end{equation} 
with $n=1,2,\dots $ the population transfer is complete. We show in Fig. \ref 
{second_order_success} the final population in state 3 as a function of $% 
A/\pi $. 
 
%%%%%%%%%%%%%%%%%%%%%%%%%%%%%%%%%%%%%%%%%%%%%%%%%%%%%%%%% 
 
%%%%%%%%%%%%%%%%%%%%%%%%%%%%%%%%%%%%%%%%%%%%%%%%%%%%%%%%% 
 
A special case of the population transfer with second-order matched pulses 
discussed in the present section is the analytical model discussed by 
Vitanov and Stenholm in \cite{Vitanov}. These authors considered a pulse 
sequence with  
\begin{equation} 
\Omega_0(t)=\frac{\alpha}{2 T}{\rm sech}^2\left(\frac{t}{T}\right),\qquad 
\theta_0(t)=\frac{\pi}{4}\left[{\rm tanh}\left(\frac{t}{T}\right)+1\right] 
\end{equation} 
and thus $\tan\theta_1 = \Omega_0(t)/2{\dot\theta}_0(t) =\alpha/\pi=$ const. 
 
%------------------------------------------------------------------------ 
 
\subsection{Population transfer via large-area third-order matched pulses} 
 
%------------------------------------------------------------------------ 
 
Next we analyze the possibility of population transfer when $\Phi 
_{3}^{\left( 3\right) }$ is exactly trapped. In order for $\Phi _{3}^{(3)}$ 
to be a constant of motion or equivalently to have third-order matched 
pulses $\theta _{2}=$const. We assume again that the system state vector $% 
\Psi $ is initially in the trapping state in which it will remain for all 
times. In order to realise population transfer from state 1 to state 2 or 3 
in this case, we furthermore must satisfy the initial conditions  
\begin{equation} 
C_{1}\left( -\infty \right) =1,\quad C_{2}\left( -\infty \right) =0,\quad 
C_{3}\left( -\infty \right) =0.  \label{atominit} 
\end{equation} 
This can be translated into a condition for the initial values of the 
dynamical phases $\theta _{0}$ and $\theta _{1}$ using Eq.(\ref{B-C}). In 
fact from  
\begin{equation} 
{\bf C}={\sf V}_{3}^{-1}{\bf B}^{(3)}={\sf V}_{3}^{-1}\left[ \matrix{0\cr 
0\cr 1}\right]  
\end{equation} 
we find  
\begin{eqnarray} 
C_{1}(t) &=&i\left( \cos \theta _{0}(t)\sin \theta _{1}(t)\sin \theta 
_{2}-\sin \theta _{0}(t)\cos \theta _{2}\right)   \label{evolut} \\ 
C_{2}(t) &=&-i\cos \theta _{1}(t)\sin \theta _{2} \\ 
C_{3}(t) &=&-i\left( \cos \theta _{0}(t)\cos \theta _{2}+\sin \theta 
_{0}(t)\sin \theta _{1}(t)\sin \theta _{2}\right) . 
\end{eqnarray} 
The initial condition it is fulfilled when  
\begin{eqnarray} 
\cos \theta _{0}\left( -\infty \right) \sin \theta _{1}\left( -\infty 
\right) \sin \theta _{2}-\cos \theta _{2}\sin \theta _{0}\left( -\infty 
\right)  &=&1, \\ 
\sin \theta _{2}\cos \theta _{1}\left( -\infty \right)  &=&0, \\ 
\cos \theta _{0}\left( -\infty \right) \cos \theta _{2}+\sin \theta _{2}\sin 
\theta _{0}\left( -\infty \right) \sin \theta _{1}\left( -\infty \right)  
&=&0. 
\end{eqnarray} 
The result is  
\begin{equation} 
\theta _{1}\left( -\infty \right) =\frac{\pi }{2},\quad \theta _{0}\left( 
-\infty \right) =\theta _{2}+\frac{\pi }{2}.  \label{initcond} 
\end{equation} 
 
From Eq.(\ref{thetan}) we find the following differential equation  
\begin{equation} 
2\frac{d\theta _{1}(t)}{dt}=\alpha \,\Omega _{1}(t),\qquad {\rm where}\qquad 
\alpha =(\tan \theta _{2})^{-1}={\rm const.} 
\end{equation} 
Introducing  
\begin{equation} 
x(t)=\tan \theta _{1}(t) 
\end{equation} 
we find furthermore  
\begin{eqnarray} 
\frac{2x\,\dot{x}}{\left( 1+x^{2}\right) ^{3/2}} &=&\alpha \Omega _{0}(t), \\ 
2\frac{d\theta _{0}(t)}{dt} &=&\frac{\Omega _{0}(t)}{x(t)}. 
\end{eqnarray} 
Integrating these equations and taking into account the initial conditions (% 
\ref{initcond}) yields  
\begin{eqnarray} 
\tan \theta _{1}(t) &=&x(t)=\sqrt{\displaystyle\frac{1}{f^{2}(t)}-1}, \\ 
\theta _{0}(t) &=&\theta _{2}+\frac{\pi }{2}+\frac{1}{\alpha }\Bigl[1-\sqrt{% 
1-f^{2}(t)}\Bigr], 
\end{eqnarray} 
where  
\begin{equation} 
f(t)=\frac{\alpha }{2}\int_{-\infty }^{t}dt^{\prime }\,\Omega _{0}\left( 
t^{\prime }\right) . 
\end{equation} 
$\Omega _{0}(t)$ is an arbitrary smooth function which we assume to vanish 
at infinity, $\Omega _{0}\left( \pm \infty \right) =0$. We still have one 
free constant $\alpha $, which we can choose. As we will show now, we can 
choose $\alpha $ such that the efficiency of the transfer from state $1$ to 
states $3$ or $2$  approaches unity. 
 
%-------------------------------------------------------------------- 
 
\subsubsection{Population transfer from ground state to state 3} 
 
In order to transfer the initial population from state $1$ to the target 
state $3$, it is necessary to satisfy the final conditions  
\begin{equation} 
\theta _{1}\left( +\infty \right) =\frac{\pi }{2},\quad \theta _{0}\left( 
+\infty \right) =\theta _{2} 
\end{equation} 
which implies 
\begin{eqnarray} 
\tan \theta _{1}(\infty ) &=&\sqrt{\frac{4}{\alpha ^{2}A^{2}}-1}\to \infty , 
\\ 
\theta _{0}(\infty ) &=&\theta _{2}+\frac{\pi }{2}+\frac{1}{\alpha }\Bigl[1-% 
\sqrt{1-\alpha ^{2}A^{2}/4}\Bigr]=\theta _{2}, 
\end{eqnarray} 
where  
\begin{equation} 
A=\int_{-\infty }^{\infty }\!\!dt\,\Omega _{0}(t) 
\end{equation} 
is the pulse area. From these condition one finds the constraint  
\begin{equation} 
\alpha =-\frac{4\pi }{\pi ^{2}+A^{2}},\qquad A\gg \pi .  \label{alfa1} 
\end{equation} 
The diabatic losses in the limit $A\gg 1$ are  
\begin{equation} 
1-\Bigl|C_{3}(\infty )\Bigr|^{2}\approx \frac{4\pi ^{2}}{2\pi ^{2}+A^{2}} 
\end{equation} 
and thus in the adiabatic limit we have essentially complete population 
transfer from state $1$ to state $3$. 
 
\begin{figure}[tbp] 
\begin{center}
\leavevmode \epsfxsize=7 true cm
\epsffile{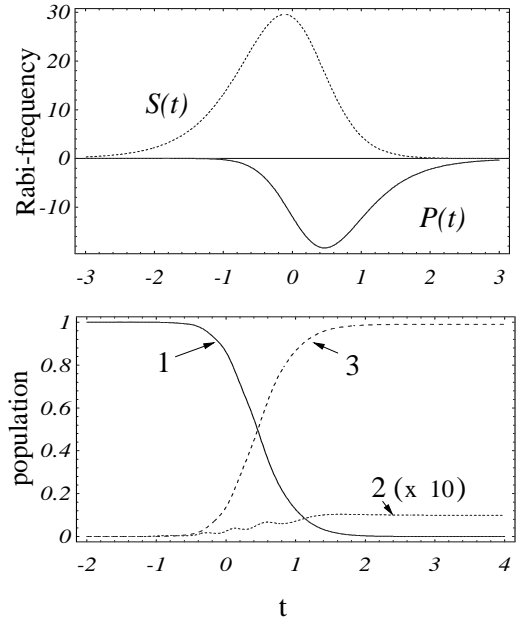}
\end{center}
\caption{Population transfer from $1$ to $3$ with third-order matched 
pulses. Upper frame shows pulses, lower frame population dynamics. Here $% 
\Omega_0(t)=A/2\, {\rm sech}^2(t)$ and $A/\pi=20$.} 
\label{third_order_lower} 
\end{figure}

Fig.\ref{third_order_lower} shows an example of population transfer with 
third-order matched pulses. Here $\Omega _{0}(t)=A/2\,{\rm sech}^{2}(t)$ and  
$A=20\pi $. Pump and Stokes pulses are shown in the upper frame and the 
population histories in the lower one. We see that the amplitudes of the 
Stokes and pump pulses are unequal. As in ordinary STIRAP the population of 
the state $2$ is small during the evolution. 
 
%%%%%%%%%%%%%%%%%%%%%%%%%%%%%%%%%%%%%%%%%%%%%%%%%%%%%%%%% 

%%%%%%%%%%%%%%%%%%%%%%%%%%%%%%%%%%%%%%%%%%%%%%%%%%%%%%%%% 
 
%------------------------------------------------------------------------ 
 
\subsubsection{Population transfer from ground state to the state 2} 
 
In order to transfer the initial population from state $1$ to state $2$, it 
is necessary to satisfy the conditions  
\begin{equation}  \label{cond2} 
\theta _1\left( +\infty \right) =0 , \quad \theta _2=\frac{ \pi }{2} 
\end{equation} 
In this case we have to fix $\alpha $ to be  
\begin{equation} 
\alpha =\frac{ 2}{A},\qquad A \gg 1. 
\end{equation} 
 
\begin{figure}[tbp] 
\begin{center}
\leavevmode \epsfxsize=7 true cm
\epsffile{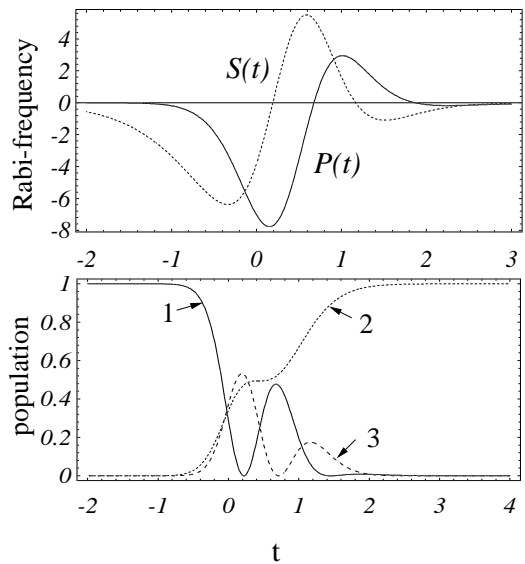}
\end{center}
\caption{Population transfer from $1$ to $2$ with third-order matched 
pulses. Upper frame shows pulses, lower frame population dynamics. Here $% 
\Omega_0(t)=A/2\, {\rm sech}^2(t)$ and $A=16$.} 
\label{third_order_upper} 
\end{figure}

Fig.\ref{third_order_upper} shows the pulses $P(t)$ and $S(t)$ and the 
evolution of the atomic populations. Here $\Omega _{0}(t)=A/2\,{\rm sech}% 
^{2}(t)$ and $A=16$. We see that the Stokes and pump pulses are in a 
counterintuitive sequence. At first the atomic population oscillates between 
state $1$ and $3$, but as the pulse sequence proceeds the whole population 
is transfered into state $2$. In other words, during the full pulse sequence 
there occur several STIRAP transitions, but due to the large nonadiabatic 
coupling the population accumulates in state $2$. 
 
%%%%%%%%%%%%%%%%%%%%%%%%%%%%%%%%%%%%%%%%%%%%%%%%%%%%%%%%% 

%%%%%%%%%%%%%%%%%%%%%%%%%%%%%%%%%%%%%%%%%%%%%%%%%%%%%%%%% 
 
%%%%%%%%%%%%%%%%%%%%%%%%%%%%%%%%%%%%%%%%%%%%%%%%%%%%%%%%%%%%%%%%%%%%%%%% 
 
\section{summary} 
 
%%%%%%%%%%%%%%%%%%%%%%%%%%%%%%%%%%%%%%%%%%%%%%%%%%%%%%%%%%%%%%%%%%%%%%% 
 
We have introduced the concept of generalized dressed states in order to 
explain the success of population transfer in stimulated Raman adiabatic 
passage with a loop coupling. If the interaction of a three-level system 
with a pair of time-dependent pump and Stokes pulses is described in terms 
of the so-called dark and bright states instead of the instantaneous 
eigenstates of the Hamiltonian, the original three-state--two-field system 
is transformed  into a system of three states coupled by two effective 
interactions \cite{Cohen_Tannoudji,Fleischhauer96}. This allows for an 
iteration procedure leading to higher-order adiabatic basis sets \cite{super}% 
. We showed that in the case of loop-STIRAP there is a higher-order trapping 
state, which is an approximate constant of motion even when the usual 
adiabaticity condition is not fulfilled. This state adiabatically rotates 
from the initial to the target quantum state of the atom and thus leads to 
efficient population transfer, however, at the expense of placing some 
population into the decaying atomic state.  
 
The concept of generalized trapping states allows the construction of pulse 
sequences which lead to an optimum population or coherence transfer also for 
small pulse areas and allows for solutions for the atomic dynamics. If pump 
and Stokes pulses fulfill certain conditions (so-called generalized matched 
pulses) the effective $3\times 3$ coupling matrix factorizes at a specific 
point of the iteration. The trapping state of the corresponding $n$th-order 
adiabatic basis is then an exact constant of motion. In this case the atomic 
dynamics reduces to a two-level problem with a real coupling which can be 
solved analytically. 
 
For ordinary matched pulses, i.e. if pump and Stokes have the same shape, 
the atomic dynamics is rather limited. The corresponding dark state is a 
constant superposition of states $1$ and $3$. In the case of generalized 
matched pulses, however, the trapping state has a time-dependent overlap 
with the bare atomic states and thus  population or coherence transfer is 
possible. We have discussed with specific example population transfer with 
second and third-order matched pulses. We found that for certain values of 
the pulse areas complete population or coherence transfer is possible. In 
the general case the diabatic losses scale non-exponentially with the 
inverse pulse area. 
 
\section*{acknowledgements} 
 
The work of RU is supported by the Alexander von Humboldt Foundation. BWS 
thanks the Alexander von Humboldt Stiftung for a Research Award; his work is 
supported in part under the auspices of the U.S. Department of Energy at 
Lawrence Livermore National Laboratory under contract W-7405-Eng-48. Partial 
support by the EU Network ERB-CHR-XCT-94-0603 is also acknowledged. 
 
%%%%%%%%%%%%%%%%%%%%%%%%%%%%%%%%%%%%%%%%%%%%%%%%%%%%%%%%%%%%%%%%%%%%%% 
 
%%%%%%%%%%%%%%%%%%%%%%%  begin references %%%%%%%%%%%%%%%%%%%%%%%%%%%%%% 


\frenchspacing 
 
\begin{references} 
\bibitem{Oreg}  J. Oreg, F. T. Hioe, and J. H. Eberly, Phys. Rev. {\bf A} 
29, 690 (1984). 
 
\bibitem{STIRAP}  U. Gaubatz, P. Rudecki, M. Becker, S. Schiemann, M. 
K\"{u}lz, and K. Bergmann, Chem. Phys. Lett. {\bf 149}, 463 (1988); for 
reviews on the experimental developments of STIRAP see: K. Bergmann and B. 
W. Shore, in {\it Molecular Dynamics and Spectroscopy by Stimulated Emission 
Pumping}, ed. by H.L. Dai and R.W. Field (World Scientific, Singapore, 1995) 
pp. 315-373;
 
\bibitem{STIRAP2}  K. Bergmann, H. Theuer, and B. W. Shore, {\it ``Coherent 
Population Transfer Among Quantum States of Atoms and Molecules''}
 Rev. Mod. Phys. 
{\bf 70},1003 (1998).  
 
\bibitem{Arimondo}  for a review on coherent population trapping see: E. 
Arimondo, Prog. in Optics {\bf 35}, 259 (1996). 
 
\bibitem{Elk95}  M. Elk, Phys. Rev. A {\bf 52}, 4017 (1995). 
 
\bibitem{Stenholm96}  T.A. Laine and S. Stenholm, Phys. Rev. A {\bf 53}, 
2501 (1996). 
 
\bibitem{Fleischhauer96}  M. Fleischhauer and A.S. Manka, Phys. Rev. A {\bf % 
54}, 794 (1996). 
 
\bibitem{Pellizzari95}  A.S. Parkins, P. Marte, P. Zoller, and H.J. Kimble, 
Phys. Rev. Lett. {\bf 71}, 3095 (1993); T. Pellizzari, S.A. Gardiner, J.I. 
Cirac, and P. Zoller, Phys. Rev. Lett. {\bf 75}, 3788 (1995). 
 
\bibitem{cavity_decay}  In adiabatic population transfer involving atoms 
coupled to a common cavity mode as used in \cite{Pellizzari95}, the transfer 
time needs to be much shorter than the cavity decay time. 
 
\bibitem{loop}  R.G. Unanyan, L.P. Yatsenko, K. Bergmann, and B.W. Shore, 
Opt. Comm. {\bf 139}, 48 (1997). 
 
\bibitem{super}  M.V. Berry, Proc. R. Soc. A {\bf 414}, 31 (1987); {\it ibid} 
{\bf 429}, 61 (1990); R. Lim and M.V. Berry, J. Phys. A {\bf 24}, 3255 
(1991). 
 
\bibitem{matched}  S.E. Harris, Phys. Rev. Lett. {\bf 72}, 52 (1994); J.H. 
Eberly, M.L. Pons, and H.R. Haq, Phys. Rev. Lett. {\bf 72}, 56 (1994). 
 
\bibitem{Malinovsky}  V.S. Malinovsky and D.J. Tannor, Phys. Rev. A {\bf 56}% 
 , 4929 (1997).
 
\bibitem{Vitanov}  N.V. Vitanov and S. Stenholm, Phys. Rev. A {\bf 55}, 648 
(1997).
 
\bibitem{Cohen_Tannoudji} A. Aspect, E. Arimondo, R. Kaiser, N. Vansteenkiste,
and C. Cohen-Tannoudji, J. Opt. Soc. Am. {\bf 6}, 2112 (1989).



\end{references}
\end{document}